\documentclass[12pt]{article}
\usepackage{amsmath}
\usepackage{amstext,amssymb,amsfonts}
\usepackage[dvips]{graphicx}
\usepackage{color}
\advance\voffset by -1.5cm
\advance\hoffset by -1.25cm
\def\be{\begin{equation}}
\def\ee{\end{equation}}
\def\ba{\begin{eqnarray}}
\def\ea{\end{eqnarray}}
\newcommand{\A}{{\cal A}}

\renewcommand{\H}{{\cal H}}

\newcommand{\Cop}{\mathbb{C}}

\begin{document}

\vspace*{-1.5cm}
\thispagestyle{empty}
\begin{flushright}
\end{flushright}
\vspace*{2.5cm}

\begin{center}
{\Large 
{\bf Constraints on extremal self-dual CFTs}}
\vspace{2.5cm}

{\large Matthias R.\ Gaberdiel$^{*}$}%
\footnotetext{$^{*}${\tt E-mail: gaberdiel@itp.phys.ethz.ch}} 
\vspace*{0.5cm}

Institut f{\"u}r Theoretische Physik, \\
ETH Z{\"u}rich\\
CH-8093 Z{\"u}rich, Switzerland\\
\vspace{2cm}
\begin{center}
July 2007
\end{center}

\vspace*{3cm}

{\bf Abstract}
\end{center}
We argue that the existence of a modular differential equation implies
that a certain vector vanishes in Zhu's $C_2$ quotient space, and we
check this assertion in numerous examples. If this connection is true
in general, it would imply that the recently conjectured extremal 
self-dual conformal field theories at $c=24 k$ cannot exist for 
$k\geq 42$.


\newpage
\renewcommand{\theequation}{\arabic{section}.\arabic{equation}}


\section{Introduction}
\setcounter{equation}{0}

Recently Witten has conjectured \cite{Witten:2007kt}, based on an
analysis of pure gravity in $AdS_3$, that a family of extremal
self-dual meromorphic bosonic conformal field theories at $c=24 k$
(with $k=1,2,\ldots$) exists. Here extremal means that, up to level
$k+1$ above the vacuum, the theory only contains the Virasoro
descendants of the vacuum state. A meromorphic conformal field theory 
(see \cite{pg} for an introduction) is self-dual if its only 
representation is the vacuum representation itself. In particular,
this implies that the vacuum character has to be invariant under the
$S$-modular transformation. Taken together, these requirements then
determine the vacuum character of this meromorphic conformal field
theory uniquely. 

For $k=1$ the meromorphic conformal field theory in question is the
famous Monster theory \cite{FLM,Borcherds} (for a beautiful
introduction see \cite{Gannon}). But for $k\geq 2$ an explicit
realisation of these theories is so far not known. The proposed
conformal field theories satisfy however a few consistency
conditions. First of all, the coefficients of the $q$-expansion are
positive integers, and thus can be interpreted as vacuum characters
of conformal field  theories. Witten also showed that the $k=2$ vacuum
amplitudes are well-defined on higher genus Riemann surfaces
\cite{Witten:2007kt}; more recently, the genus $2$ amplitude of the
$k=3$ theory was shown to be consistent (by some other methods)
\cite{Gaiotto:2007xh}. Their method determines also the genus 2
partition functions uniquely up to $k\leq 10$.  
\smallskip

While these are impressive consistency checks, they essentially only
test the modular properties of vacuum expectation values, and are thus
not very sensitive to the inner workings of the theory. (For example,
on the level of the torus amplitudes, one would expect that there are 
infinitely many self-dual conformal field theories at $c=24$, since 
one may add to $J(q)=j(q)-744$ any positive integer. On the other
hand, it is believed that there are only $71$ such theories 
\cite{Schellekens:1992db}.) It would thus be very desirable to subject
these theories to consistency conditions that go beyond these
considerations. In this paper we want to make one small step in this
direction by analysing the structure of the modular differential
equation for these theories.  
\smallskip

It has long been known that all the characters of a rational conformal
field theory satisfy a common modular differential equation
\cite{MMS,MMS1}. From a mathematical point of view, this differential
equation can be obtained quite generally for theories that satisfy the
so-called $C_2$ condition \cite{Zhu}; this is believed to be the case
for all rational conformal field theories. For theories satisfying
this condition, there exists an integer $s$ for which $L_{-2}^s\Omega$
plus some correction terms of lower conformal weight lie in a certain 
subspace $O_q(\H_0)$ --- for precise definitions see section~2. This
then leads to a modular differential equation of order $s$. On the
other hand, such a relation in $O_q(\H_0)$ can only exist if  
$L_{-2}^{s}\Omega \in O_{[2]}$. In turn this requires that the vacuum
representation possesses a null-vector at conformal weight $2s$.  

It seems very natural to believe (and we shall show that this is at
least true in many examples) that every modular differential equation
arises in this fashion. Thus if the characters of the chiral algebra
satisfy a modular differential equation of order $s$, this suggests
that $L_{-2}^s\Omega\in O_{[2]}$, which in turn implies that the 
vacuum representation has to have a null vector at level $2s$. Applied
to the above candidate theories we find that for  $k\geq 42$, this
predicts the existence of a null vector at a level 
less than $k+1$. On the other hand, since the theory up to level $k+1$
is just the Virasoro theory at $c=24 k$, we know that no such null
vector exists. Thus our analysis suggests that at least the theories
with $k\geq 42$ are inconsistent.
\bigskip

The paper is organised as follows. In the next section we review
the modular differential equation from the point of view of
\cite{Zhu}. In section~3 we explain why we expect the order of the
modular differential equation to be related to the property that
$L_{-2}^s\Omega\in O_{[2]}$. We also check this claim
explicitly for a number of theories, in particular, the minimal
models, the su(2) affine theories, su(3) at level $k=1,2$ and the
self-dual theories corresponding to $e_8$ (at level $k=1$), 
$e_8\oplus e_8$, $e_8\oplus e_8 \oplus e_8$ as well as the Monster
theory. In section~4 we then apply this technique to the proposed 
self-dual meromorphic conformal field theories at $c=24 k$, and find
that the corresponding null vector seems to arise at levels that are
too low (for $k\geq 42$).

\section{The modular differential equation}
\setcounter{equation}{0}

Let us begin by reviewing the definition of the $C_2$ criterion of Zhu
\cite{Zhu}. We denote the vacuum  representation of the chiral algebra
by $\H_0$, and the fields of the chiral algebra that generate $\H_0$
from the vacuum state $\Omega$ by $S^i$; the conformal weight of $S^i$
is $h_i$, where $h_i$ is a positive integer. In the following we shall
only consider bosonic bosonic conformal field theories, although the
main ideas will also apply to fermionic theories. We shall also
assume that the spectrum of $L_0$ is bounded from below by zero, and
that there is a unique vector with $L_0\Omega = 0$, the vacuum
$\Omega$. 

With these preparations we now define the subspace $O_{[2]}$ of $\H_0$
(we are using the same notation as in \cite{Gaberdiel:2000qn}) as the
vector space that is spanned by the vectors of the form
\be
S^i_{-h_i-1} \phi \ , \qquad \hbox{where} \quad 
\phi \in \H_0 \ .
\ee
A chiral algebra (or vertex operator algebra) satisfies the
$C_2$-criterion, if $O_{[2]}$ has finite codimension in $\H_0$, 
{\it i.e.}\ if the quotient space 
$A_{[2]}=\H_0 / O_{[2]}$ is finite-dimensional. It was conjectured by
Zhu \cite{Zhu} that all rational conformal field theories satisfy the
$C_2$ criterion. This has also been confirmed in numerous cases. 
Obviously, if the chiral algebra satisfies the $C_2$-criterion, there 
exists a positive integer $s_0$ such that 
\be\label{O2s}
L_{-2}^{s_0}\Omega \in O_{[2]} \ .
\ee
\smallskip

For the discussion of the characters (or torus amplitudes) a
different, but closely related quotient space is of relevance. To
define it, we consider the ring of modular forms 
$\Cop[E_4(q),E_6(q)]$ that is generated by the Eisenstein series
$E_4(q)$ and $E_6(q)$. Recall that a modular form of weight $k$ is a
function $f(\tau)$ satisfying 
\be
f\left( \frac{a\tau + b}{c\tau + d}\right) = (c\tau + d)^{k} f(\tau) 
\ ,
\ee
where as always $q=e^{2\pi i \tau}$. We furthermore require that
$f(\tau)$ has a Taylor series expansion in non-negative integer powers
of $q$. The Eisenstein series $E_4$ and $E_6$ are modular forms of
weight $4$ and $6$, respectively, and they freely generate 
the ring of all modular forms. Our conventions for the Eisenstein
series are 
\begin{eqnarray}
E_2(q) & = & 1-24\, q-72\,  q^2-96\,  q^3-168\,  q^4
-144\,  q^5-288\, q^6 - \cdots
\ , \nonumber \\
E_4(q) & = & 1 + 240 \, q + 2160\,  q^2 + 6720\,  q^3 
+ 17520\,  q^4 + 30240\,  q^5 + 60480\,  q^6 + \cdots \ , 
\nonumber \\
E_6(q) & = & 1 - 504\,  q - 16632\,  q^2 - 122976\,  q^3 
- 532728\,  q^4 - 1575504\,  q^5 - 4058208  q^6 - \cdots .  
\nonumber 
\end{eqnarray}
The Eisenstein series $E_2(q)$ is not a modular form since it has a
conformal anomaly
\be
E_2\left( \frac{a\tau + b}{c\tau + d}\right) = (c\tau + d)^{2} 
\left( E_2(\tau) + \frac{6}{i \pi} \, \frac{c}{c\tau +d} \right) \ . 
\ee
It will also play an important role in the following.

With these preparations we now consider the `module'
$\H_0[E_4(q),E_6(q)]$ of the vacuum representation $\H_0$ over
$\Cop[E_4(q),E_6(q)]$; this consists of linear combinations of vectors  
in $\H_0$, where the coefficients are polynomials in $E_4(q)$ and
$E_6(q)$. We then define the subspace $O_q(\H_0)$ of
$\H_0[E_4(q),E_6(q)]$ to be generated by the vectors of the form 
\be
S^i_{[-h_1-1]} \phi + 2
\sum_{k=2}^\infty  (2k-1) \,\zeta(2k) \,E_{2k}(q) \,
S^i_{[2k-h_i-1]} \phi \qquad \phi \in \H_0 \ .
\ee
Here $E_{2k}$ for $k\geq 4$ are the higher Eisenstein series that
can be written in terms of polynomials in $E_4$ and $E_6$.
The modes $S_{[n]}$ are the natural modes on the torus and can be
expressed in terms of the original modes as (see \cite[(4.2.3)]{Zhu}
for an explicit formula)
\be
S^i_{[n]} = S^i_n + \sum_{m\geq 1} c_{m,n}(h_i) S^i_{n+m} \ ,
\ee
where $c_{m,n}(h^i)$ are constants.
\smallskip

The motivation for the definition of $O_q(\H_0)$ comes from the fact
that if $\psi\in O_q(\H_0)$ then 
\be
{\rm Tr}_{\H_j} \left( V_0(\psi) q^{L_0 - \frac{c}{24}} \right) = 0  
\ , 
\ee
where $\H_j$ is an arbitrary representation of the chiral conformal
field theory. Put differently, $O_q(\H_0)$ describes the subspace of
$\H_0[E_4(q),E_6(q)]$ whose one-point torus amplitudes vanish (see
also \cite{Witten:1987tv}). 

It is obvious from the above definitions that one can think of 
$O_q(\H_0)$ as a `deformation' of $O_{[2]}$. It is then easy to see 
(and explained in \cite{Zhu}) that if a chiral conformal field theory
satisfies the $C_2$ criterion, then there exists a positive integer
$s$, such that  
\be\label{Virdes}
\left(L_{[-2]}^s + \sum_{r=0}^{s-1} f_r(q) L^r_{[-2]} \right)
\Omega \in O_{q}(\H_0) \ ,
\ee
where each $f_r(q)$ is polynomial in $E_4$ and $E_6$. 

The smallest such integer $s$ will be called the {\it size} of 
the chiral algebra $\A$. Because of the grading of 
$\H_0[E_4(q),E_6(q)]$ (in terms of modular weight and conformal weight
with respect to $L_{[0]}$) one can show that (\ref{Virdes}) implies
that (\ref{O2s}) holds for the same $s$\footnote{This will be
explained in \cite{GK} where a more 
comprehensive description of the whole approach will be given.}, 
{\it i.e.}\ that $L_{-2}^s\Omega\in O_{[2]}$. The converse may in
general not be true, although I do not know of any explicit
counterexample.   

If we insert the zero mode of the vector (\ref{Virdes}) into the
character of an arbitrary representation  we find that it vanishes,
given the definition of $O_q(\H_0)$. On the other hand, using standard
conformal field theory techniques, Zhu showed \cite{Zhu} that inside
the trace, the zero mode of (\ref{Virdes}) can be expressed in terms of
polynomials of $L_0$, involving as coefficients polynomials in the
Eisenstein series $E_2$, $E_4$ and $E_6$. In turn, each $L_0$ can be
expressed in terms of a derivative with respect to $q$, and one thus
obtains a differential equation of the form\footnote{It would be very 
interesting to understand the relation between this differential
equation and the one recently considered in \cite{Bantay:2005vk}.} 
\be
\left[ \left( q \frac{d}{dq} \right)^s  + \sum_{0\leq r<s} 
\hat{f}_r(q) \left( q \frac{d}{dq} \right)^r \right] \chi_j(q) = 0 \ ,
\ee
where the $\hat{f}_r(q)$ are polynomials in the Eisenstein series
$E_2$, $E_4$ and $E_6$ (that are independent of which character
$\chi_j$ is being considered). This equation can be thought of as
being the modular differential equation of \cite{MMS,MMS1} (for
earlier work see \cite{Eguchi:1986sb,Anderson:1987ge}; further
developments are described in \cite{Eholzer,ES,Flohr:2005cm}). Using
the fact that it has to transform covariantly under the modular group,
it can be brought into the form
\be\label{ansatz}
\left[ D^s + \sum_{r=0}^{s-2} f_{r}(q)\, D^{r} \right] \chi_j(q) = 0 \ , 
\ee
where each $f_r(q)$ is a polynomial in $E_4(q)$ and $E_6(q)$ of
modular weight $2(s-r)$, and 
\begin{equation}
D^r = {\it cod}_{(2r-2)}\cdots{\it cod}_{(2)} {\it cod}_{(0)}\ . 
\end{equation}
Here $cod_{s}$ is the modular covariant derivative that maps a modular
form of weight $s$ to one of weight $s+2$, 
\begin{equation}
\label{cod}
{\it cod}_{(s)} = q \frac{d}{dq}
                - \frac{s}{12}  E_2(q) \ .
\end{equation}
Note that the modular anomaly of $E_2(q)$ is crucial in order for this
to be modular covariant.

\section{The order of the differential equation}
\setcounter{equation}{0}

It is believed \cite{MMS,MMS1} that the minimal order of the
differential equation always agrees with the number of independent
characters, {\it i.e.}\ with the number of irreducible
representations of the chiral algebra, where pairs of conjugate
representations (that lead to the same character) are only counted
once. While this is true in many cases (that were checked in
\cite{MMS,MMS1}), it cannot be true for self-dual conformal field
theories. (Self-dual chiral algebras are characterised by the property
that they only possess one  representation, namely the vacuum
representation itself.) Indeed, if this was so, the minimal order of
the modular differential equation for self-dual chiral algebras would
have to be one; but then it would necessarily have to be of the form  
\be
q \frac{d}{dq}\, \chi = 0 
\ee
which only has the trivial solution, $\chi=1$. It therefore follows
that the minimal order of the modular differential equation 
{\it cannot} always agree with the number of independent
characters. In fact, the present argument shows that the minimal order
of the modular differential equation is always at least two. 

In the following we want to argue that instead the minimal order of the
modular differential equation always agrees with the size of the
chiral algebra, {\it i.e.}\ with the smallest $s$ for which
(\ref{Virdes}) holds. This circumvents the above problem since 
(\ref{Virdes}) implies that $L_{-2}^s\Omega \in O_{[2]}$, which is 
only possible for $s\geq 2$. To see this we recall that the
non-trivial fields of the theory all have $h_i\geq 1$. It is
therefore impossible to find a null-vector relation that would make
$L_{-2}\Omega$ an element of $O_{[2]}$, and hence $s=1$ in (\ref{O2s})
is never possible.        
\smallskip

One direction of this proposed correspondence is straightforwardly
proven: the argument of the previous section implies that we can
always find a modular differential equation at order $s$ if $s$ is
the size of the chiral algebra. The above statement therefore amounts
to the assertion that also the converse is true, {\it i.e.}\ if all 
characters of the conformal field theory satisfy (\ref{ansatz}) for
some $s$, that there 
exists a relation of the form (\ref{Virdes}) with the same
$s$. This should follow from the arguments of Zhu \cite{Zhu} --- see  
the more comprehensive description in \cite{GK}.

For the moment we shall not attempt to prove this, but rather give
examples that suggest the truth of this assertion. In each case we
shall show that the order of the differential equation is equal
to the smallest $s_0$ for which (\ref{O2s}) holds. In particular, this
then shows that if the theory satisfies a modular differential
equation of order $s$, then $L_{-2}^s\Omega \in O_{[2]}$, which is the
main conjecture relevant for the analysis of the extremal conformal
field theories at $c=24 k$. We begin with some simple cases for which
the analysis can be done completely.

\subsection{Minimal models}

The Virasoro minimal models arise for the central charges $c=c_{p,q}$
with 
\be
c_{p,q} = 1 - \frac{6(p-q)^2}{pq} \ ,
\ee
where $p,q\geq 2$ are coprime integers. They define rational conformal
field theories with $(p-1)(q-1)/2$ inequivalent highest weight
representations; the corresponding conformal weights are given by 
\be
h_{(r,s)} = \frac{(rp-qs)^2 - (p-q)^2}{4pq} \ ,
\ee
where $1\leq r \leq q-1$ and $1\leq s \leq p-1$, and we have the
identification $h_{r,s}=h_{q-r,p-s}$. For Virasoro minimal models the
analysis of \cite{MMS,MMS1} applies, and it follows that the order of 
the modular differential equation is precisely equal to the number of
independent characters, {\it i.e.}\ to $(p-1)(q-1)/2$.  

For Virasoro theories the space $A_{[2]}$ can be taken to be generated
by the vectors $L_{-2}^l\Omega$, where $l=0,1,\ldots$. For $c=c_{p,q}$ 
the vacuum Verma module has a null-vector at level $(p-1)(q-1)$ for
which the coefficient of $L_{-2}^{(p-1)(q-1)/2}\Omega$ does not
vanish \cite{FF}. This implies that $L_{-2}^{(p-1)(q-1)/2}\Omega$ is
in fact in $O_{[2]}$ (since it differs from vectors in $O_{[2]}$ by a
null-vector), and thus that the minimal $s_0$ in (\ref{O2s}) is indeed
$(p-1)(q-1)/2$.

\subsection{SU(2) WZW models at level $k$}

The next simple class of models for which we can give a complete
description are the su(2) current
theories at level $k$. It is well known that these chiral algebras
have $k+1$ irreducible inequivalent representations, namely those
characterised by the spin $j$ of the highest weight space with
$j=0,\tfrac{1}{2},\ldots,\tfrac{k}{2}$. Again the analysis of
\cite{MMS,MMS1} applies, and it follows that the order of the modular
differential equation is $k+1$. This was also worked out explicitly in
\cite{MMS} for the case $k=1$. 

On the other hand, the quotient space $A_{[2]}$ for these models was
analysed in \cite{GN}. For $k=1$ we found
\be
A_{[2]}\bigl(\widehat{su}(2)_1\bigr) = 
{\bf 1}_0 \oplus {\bf 3}_1 \oplus {\bf 1}_2 \ , 
\ee
where we have decomposed $A_{[2]}$ in terms of representations of the
zero modes of $\widehat{su}(2)$, and the index indicates at which
conformal weight theses states appear. [So for example, ${\bf 3}_1$
refers to the three states $J^a_{-1}\Omega$, {\it etc.}] The state at
level two is precisely $L_{-2}\Omega$, which is therefore {\it not} in 
$O_{[2]}$. On the other $L_{-2}^2\Omega=0$ in $A_{[2]}$, and hence the
minimal $s_0$ in (\ref{O2s}) is $s_0=2=k+1$, in agreement with the
above.  

\noindent For $k=2$ we find instead
\be
A_{[2]}\bigl(\widehat{su}(2)_2\bigr) = 
{\bf 1}_0 \oplus {\bf 3}_1 \oplus {\bf 1}_2 \oplus 
{\bf 5}_2 \oplus {\bf 3}_3  \oplus {\bf 1}_4 \ .
\ee
The states ${\bf 1}_2$ and ${\bf 1}_4$ correspond to $L_{-2}\Omega$
and $L_{-2}^2\Omega$, respectively, and thus only the state
$L_{-2}^3\Omega\in O_{[2]}$. Thus the minimal $s_0$ in (\ref{O2s}) is 
$s_0=3=k+1$, as expected. For $k=3$ we find 
\be
A_{[2]}\bigl(\widehat{su}(2)_3\bigr) = 
{\bf 1}_0 \oplus {\bf 3}_1 \oplus {\bf 1}_2 \oplus 
{\bf 5}_2 \oplus {\bf 3}_3 \oplus {\bf 7}_3  \oplus 
{\bf 1}_4 \oplus {\bf 5}_4 \oplus {\bf 3}_5 \oplus {\bf 1}_6 \ .
\ee
Now the singlet vectors correspond to $L_{-2}^l\Omega$ with
$l=0,1,2,3$, and hence only $L_{-2}^4\Omega\in O_{[2]}$, leading to 
the minimal $s_0=4=k+1$, again as expected. 

It is not difficult to guess now how the structure will continue for
all $k$: at every $k$, $A_{[2]}(\widehat{su}(2)_k)$ will contain the
singlet vectors ${\bf 1}_{2n}$ with $n=0,1,2,\ldots, k$,
corresponding to $L_{-2}^{n}\Omega$. Thus only 
$L_{-2}^{k+1}\Omega\in O_{[2]}$, leading to the minimal $s_0=k+1$.

\subsection{SU(3) WZW model at level $k=1,2$}

Unfortunately, the analysis of the quotient space $A_{[2]}$ becomes
increasingly complicated for affine algebras of higher rank, and we do
not know any general formulae beyond su(2). However, we can still give
the results for low levels, for example for su(3) at $k=1$ and $k=2$.

For $k=1$ there are three irreducible representations (namely the
vacuum representation, as well as those associated to ${\bf 3}$ and 
${\bf \bar{3}}$ of su(3)); since the ${\bf 3}$ and ${\bf \bar{3}}$ 
representations are conjugate representations, they lead to the same
character and we thus expect to find a second order modular
differential equation. On the other hand, the $A_{[2]}$ space consists
of \cite{GN} 
\be
A_{[2]}(\widehat{su}(3)_1) = 
{\bf 1}_0 \oplus {\bf 8}_1 \oplus {\bf 1}_2 \oplus
{\bf 8}_2 \oplus {\bf 1}_3  \ , 
\ee
and thus $L_{-2}^l\Omega$ with $l=0,1$ is in $A_{[2]}$ --- these are
the two singlet states at levels $0,2$ --- but
$L_{-2}^2\Omega\in O_{[2]}$, showing that su(3) at level $k=1$ has
indeed minimal $s_0=2$.

At level $k=2$, there are six irreducible representations 
--- in addition now also the ${\bf 6}$, ${\bf \bar{6}}$ and ${\bf 8}$ 
appear --- but two of them are complex conjugates of one another, and
hence we expect a fourth order differential equation. On the other
hand, the $A_{[2]}$ space consists of \cite{GN}  
\begin{eqnarray}
A_{[2]}(\widehat{su}(3)_2) & = & 
{\bf 1}_0 \oplus {\bf 8}_1 \oplus 
{\bf 1}_2 \oplus {\bf 8}_2 \oplus {\bf 27}_2 \oplus 
{\bf 1}_3 \oplus {\bf 10}_3 \oplus \overline{\bf 10}_3 \oplus
{\bf 8}_3 \oplus {\bf 27}_3 
\nonumber \\
& & \oplus 
{\bf 1}_4 \oplus 2 \cdot {\bf 8}_4 \oplus {\bf 27}_4  
\oplus {\bf 1}_5 \oplus {\bf 8}_5 \oplus {\bf 1}_6  \ . 
\end{eqnarray}
This is in agreement with the expectation that 
$L_{-2}^4\Omega\in O_{[2]}$, but that no smaller power of $L_{-2}$
satisfies this condition. 
\medskip

So far we have discussed rational conformal field theories for which
the order of the modular differential equation coincides precisely
with the number of independent characters, as suggested in
\cite{MMS,MMS1}. Incidentally, these are also the theories for
which Zhu's algebra $A_{[1,1]}$ \cite{Zhu} has the same dimension as
$A_{[2]}$ \cite{GN}, see also 
\cite{Gaberdiel:1998fs,Gaberdiel:1999mc}.\footnote{The question of 
when these dimensions agree or disagree was in fact the motivation for 
determining the structure of $A_{[2]}$ for the various examples in
\cite{GN}. An introduction to Zhu's algebra can be found in
\cite{Gaberdiel:1998fs,Gaberdiel:1999mc}.} However, one would expect
potential problems with our 
proposal to arise for those cases where 
$\dim A_{[1,1]} < \dim A_{[2]}$. In particular, this happens for the 
self-dual theories since they have $\dim A_{[1,1]}=1$, whereas
$\dim A_{[2]} \geq 2$ because $A_{[2]}$ always contains at least the 
vacuum as well as $L_{-2}\Omega$.

\subsection{The self-dual $e_8$ level $1$ theory.}

The simplest self-dual conformal field theory (and conjecturally the
only self-dual conformal field theory at $c=8$ --- see
\cite{pg,Hoehn}) is the $e_8$ affine theory at $k=1$. The vacuum
character of this theory is simply  
\be
\chi_{e_8}(q) = j(q)^{1/3} = q^{-1/3} \left(
1 + 248\, q + 4124 \, q^2 + 34752\, q^3 + 213126\, q^4 
+ 1057504\, q^5 + \cdots
\right) \ .
\ee
Since it is the only character of this chiral algebra, we can
systematically search for the differential equation of the
type (\ref{ansatz}) of smallest order that annihilates
$\chi_{e_8}(q)$. One easily finds 
that 
\be\label{e8d}
\left[D^2 - \frac{1}{6} E_4(q) \right] \chi_{e_8}(q) = 0 \ .
\ee
On the other hand, the quotient space $A_{[2]}=\H_0 / O_{[2]}$ for
this theory was determined explicitly in \cite{GN}, where it was found
to consist of the $e_8$ representations 
\be
A_{[2]} = {\bf 1}_{0} \oplus {\bf 248}_1 \oplus 
{\bf 3875}_{2} \oplus {\bf 1}_{2} \ , 
\ee
and the index denotes again the conformal weight at which
these states appear. To obtain this result we used that the complete
null-space of the $e_8$ level $k=1$ theory is generated from the null
vectors that lie in the {\bf 27000} representation of $e_8$ at
conformal weight $h=2$. Furthermore, we may take $A_{[2]}$ to be
generated by the completely symmetric powers of the `adjoint'
generators $J^a_{-1}$ acting on the vacuum, since any commutator term
will automatically lie in $O_{[2]}$. To determine $A_{[2]}$ we
therefore have to determine the intersection of descendants of the
states in the {\bf 27000} with symmetrised tensor products of 
$J^a_{-1}$ acting on the vacuum. This can be analysed per computer
(using LiE and a C-program). 

The only singlet at level $2$ is the state
$L_{-2} \Omega$, and since it survives in $A_{[2]}$, it is not in
$O_{[2]}$; on the other hand, $A_{[2]}$ does not contain any states at
level $4$, and thus $L_{-2}^2\Omega\in O_{[2]}$. Thus it follows that
the $e_8$ level $k=1$ theory has indeed minimal $s_0=2$, in agreement
with the order of the modular differential equation (\ref{e8d}). 

In contrast to the situation discussed in \cite{MMS,MMS1}, the modular
differential equation has higher order (namely two) than the number of
independent characters (which is one here). One may therefore ask
what the other solutions of the modular differential equation
correspond to. For the case at hand, there is only one additional
solution (in addition to $\chi_{e_8}(q)$), which is of the form
\be
\hat\chi(q) = q^{\frac{1}{2}} \left(
1 + \frac{228}{11} q + \frac{34938}{187} q^2 + 
\frac{5163352}{4301} q^3 + \cdots \right) \ . 
\ee
This does not seem to correspond to a character of any representation;
in particular, the coefficients in the $q$-expansion do not seem to be
integers.\footnote{I have determined them up to order $q^{11}$, and the
denominator continues to contain new prime-factors as one increases
the order of the expansion.} I suspect that the $e_8$ level $k=1$
theory has a second (independent) modular differential equation at
third order (which comes from the fact that also
$L_{-2}^3\Omega \in O_{[2]}$), and that only $\chi_{e_8}$ (but 
not $\hat\chi(q)$) is a solution 
to both differential equations. (It is not difficult to find such a
third order modular differential equation, but it is not uniquely
determined by this constraint.) I suspect that the same will also
happen for the other self-dual examples that we are about to discuss.

\subsection{The self-dual $e_8\oplus e_8$ theory}

The next simplest self-dual conformal field theory is the tensor
product of two such theories at $c=16$. The corresponding character is
\be
\chi_{e_8\oplus e_8}(q) = j(q)^{2/3} = q^{-2/3} \left(
1 + 496\, q + 69752 \, q^2 + 2115008\, q^3 + 34670620 \, q^4 + 
\cdots
\right) \ .
\ee
Again,  we can systematically search for the differential equation of
the type (\ref{ansatz}) of smallest order that annihilates
$\chi_{e_8\oplus e_8}(q)$, and we find\footnote{The two coefficients
in front of the two terms $E_6$ and $E_4 D$ are not uniquely
determined by this condition; we have given the result for the case
that the $E_4 D$ term is absent.}
that 
\be\label{e8e8d}
\left[D^3 + \frac{5}{9} E_6(q) \right] \chi_{e_8\oplus e_8}(q) = 0 \ .
\ee
Thus we expect that this chiral algebra has minimal $s_0=3$. If we 
denote the Virasoro generators of the two $e_8$ copies by $L^{(1)}$ and
$L^{(2)}$, then we know from the previous analysis that 
$L^{(j)}_{-2} \Omega \not\in O_{[2]}$, but 
$L^{(j) \, 2}_{-2}\, \Omega \in O_{[2]}$. Since
$L=L^{(1)}+L^{(2)}$, we find that 
\begin{eqnarray}
L_{-2} \Omega & = & 
L^{(1)}_{-2} \Omega  + L^{(2)}_{-2} \Omega \not\in O_{[2]}  \nonumber
\\
L_{-2}^2 \Omega & = & 
L^{(1) \, 2}_{-2}\,  \Omega  + 
L^{(2) \, 2}_{-2}\,  \Omega 
+ 2 L^{(1)}_{-2} L^{(2)}_{-2}  \Omega  \not\in O_{[2]} \label{mixed}
\\
L_{-2}^3 \Omega & = & 
L^{(1) \, 3}_{-2}\,  \Omega  
+ 3 L^{(1) \, 2}_{-2} \, L^{(2)}_{-2}  \Omega 
+ 3 L^{(1) }_{-2} L^{(2)\, 2}_{-2} \, \Omega 
+ L^{(2) \, 3}_{-2} \, \Omega   \in O_{[2]} \nonumber
\end{eqnarray}
in agreement with the above expectation. Note that the last term in
(\ref{mixed}) is not in $O_{[2]}$ since 
$L^{(1)}_{-2} L^{(2)}_{-2} \Omega \not\in O_{[2]}$, but that the
expression in the third line is in $O_{[2]}$ since 
$L^{(j) \, 2}_{-2}\, \Omega \in O_{[2]}$. We are using here that 
$O_{[2]}(\H_1\oplus\H_2)=O_{[2]}(\H_1) \oplus O_{[2]}(\H_2)$, as is 
obvious from the definition of the $O_{[2]}$ space.

\subsection{The Monster theory}

The most interesting self-dual conformal field theory is probably the
Monster conformal field theory constructed by Frenkel, Lepowsky and
Meurman \cite{FLM} (see also \cite{Borcherds}). Its partition function
is the well known $J(q)=j(q)-744$ function
\be\label{Mchar}
\chi_{M} = q^{-1} \left( 
1 + 196884 \, q + 21493760\, q^2 + 864299970\, q^3 + 
20245856256\, q^4 + \cdots \right) \ .
\ee
One again finds that it satisfies a third order modular differential
equation which is now uniquely determined by this constraint to be 
\be\label{Monstd}
\left[D^3 + \frac{16}{31} E_6(q) 
- \frac{290}{279} E_4(q) \, D \right] \chi_{M}(q) = 0 \ . 
\ee
Thus we again expect the Monster theory to have minimal $s_0=3$. This
can also be independently confirmed. As we mentioned before, for any 
non-trivial theory $L_{-2}\Omega \not\in O_{[2]}$. 
Since the Monster theory does not contain any fields with $h=1$, 
$O_{[2]}$ does not contain any states at level $4$ since any
such state would have to be of the form $S_{-3} \phi$ for some $S$ of
$h=2$, and with $\phi$ at level one. But since the Monster theory does
not contain any states at level one, no such element of $O_{[2]}$ can
exist. At level six, on the other hand, this argument breaks down. At
level six we have singlet states with respect to the Monster group
that arise from Virasoro descendants of the vacuum, namely 
\be\label{Vid}
L_{-2}^3 \Omega\ , \quad 
L_{-3}^2 \Omega\ , \quad
L_{-4} \, L_{-2}  \Omega \ , \quad
L_{-6} \Omega \ .
\ee
In addition we can get three singlet states of the form (see
also \cite{M,T}) 
\be\label{Wdes}
\sum_{ij} c_{ij} L_{-2} W^i_{-2} W^j_{-2} \Omega \ , \qquad
\sum_{ij} c_{ij} W^i_{-3} W^j_{-3} \Omega  \ , \qquad
\sum_{ij} c_{ij} W^i_{-4} W^j_{-2} \Omega
\ee
where the $W^i$ denote the 196883 fields of conformal weight $h=2$
that transform in an irreducible Monster representation, and the
$c_{ij}$ are the coefficients that pick out the trivial Monster
representation in this tensor product. Finally, we have the singlet
state  
\be\label{Vdes}
\sum_{\alpha\beta} d_{\alpha\beta} 
\hat{W}^\alpha_{-3} \hat{W}^\beta_{-3} \Omega \ , 
\ee
where the $\hat{W}^\alpha$ denote the $21296876$ Virasoro primary
fields of conformal weight $h=3$ that also transform in an irreducible
Monster representation; the constants $d_{\alpha\beta}$ are again the 
appropriate Clebsch-Gordon coefficients. 

Of these eight states, only the first states in (\ref{Vid}) and 
(\ref{Wdes}), and the state in (\ref{Vdes}) do not lie manifestly in
$O_{[2]}$. Now it is known from the decomposition of coefficients in
(\ref{Mchar}) (see for example \cite{McKS}) that at level $6$, there
are only as many Monster invariant states as there are Virasoro 
descendants\footnote{This was already noted in 
\cite{DM,Hoehn,Craps:2002rw}.},
namely four. Thus we must have four linear relations between these
eight states. Unless there are some unexpected
cancellations, we therefore expect that these relations allow us to
rewrite $L_{-2}^3 \Omega$ in terms of elements in $O_{[2]}$. Thus  we
expect that $L_{-2}^3 \Omega\in O_{[2]}$, {\it i.e.} that the minimal
$s_0$ in (\ref{O2s}) is $s_0=3$ for the Monster theory.

\subsection{The other self-dual theories at $c=24$}

At $c=24$ there are a number of other self-dual theories; these
include, in particular, the lattice theories and their orbifolds 
\cite{Dolan:1989kf,Dolan:1994st}. (A complete list has been
conjectured in \cite{Schellekens:1992db}.) 

Their characters differ from that of
the Monster by a constant $K$, $\chi(q) = J(q) + K$. Provided that
$K\neq 744$, $\chi(q)$ also satisfies a third order differential
equation of the same type as (\ref{Monstd}). Unfortunately, we cannot
directly calculate the $A_{[2]}$ spaces for these examples, and we
cannot therefore compare this to the minimal value of $s_0$ in
(\ref{O2s}). However, the case $K=744$ is interesting, since this
describes precisely the character that occurs for the 
$e_8\oplus e_8 \oplus e_8$ theory,  
\be
\chi_{e_8\oplus e_8 \oplus e_8}(q) = J(q) + 744 = j(q) \ . 
\ee
In this case, $\chi_{e_8\oplus e_8 \oplus e_8}(q)$ does not satisfy a
third order equation, but only a fourth order equation. This is in
perfect agreement with the fact that only $L_{-2}^4\Omega\in O_{[2]}$
since  
\be
L_{-2}^3\Omega = 
6\, L_{-2}^{(1)} \, L_{-2}^{(2)}\, L_{-2}^{(3)} \, \Omega
+ v \not\in O_{[2]} \ , \qquad v\in O_{[2]} \ ,
\ee
where $L_n^{(i)}$, $i=1,2,3$ are the Virasoro modes of the
$i^{\text{th}}$ $e_8$ theory. The same argument implies also that 
the theory $e_8^{\oplus l}$ has minimal $s_0=l+1$. At least for the
first few $l$ this agrees with the order of the corresponding modular 
differential equation.\footnote{I thank the referee for drawing my
attention to this point.}

\section{Extremal self-dual CFTs at $c=24 k$}
\setcounter{equation}{0}

We now want to apply these ideas to the extremal self-dual conformal
field theories at $c=24 k$ that were recently proposed by Witten
\cite{Witten:2007kt}. As is explained there, the condition that the
theories are extremal and self-dual determines their partition
function uniquely. The assumption of extremality means that the vacuum
character is of the form 
\be
\chi_k = q^{-k} \left( 
\prod_{n=2}^{\infty} \frac{1}{1-q^n} + {\cal O}(q^{k+1}) \right) \ .
\ee
In terms of the representation space $\H_0$ this means that, up to
level $k$, it is generated from the vacuum by the Virasoro modes. Thus
the fields of the chiral algebra contain, apart from the stress energy
tensor with $h=2$, only primary fields with $h_i\geq k+1$. 

As before for the examples in section 3.4 -- 3.6 we can now determine
the lowest order modular differential equation that annihilates
$\chi_k$. Since the modular differential equation is covariant with
respect to the modular group, we are looking for an equation of the
form  
\be\label{ansatz1}
\left[ D^s + \sum_{r=0}^{s-2} f_r(q)\, D^r \right] \chi_k(q) = 0 \ .
\ee
We want to determine the minimal value of $s$ for which such an
equation exists. Let us denote by $h(s)$ the number of monomials in
$E_4$ and $E_6$ that have total modular weight $2s$. The total number
of free parameters in a modular differential equation of order $s$ is
then 
\be
p(s) = \sum_{t=2}^{s} h(t) 
\ee
since each $f_r(q)$ in (\ref{ansatz1}) is a modular form of weight
$2(s-r)$.  The first few values of $p(s)$ (together with the
corresponding values of $h(s)$) are tabulated in table~1. It is easy
to see that asymptotically  $h(s) \sim C_1\, s$ and 
$p(s)\sim C_2\, s^2$.   

\begin{table}[hbt]
\begin{center}
\begin{tabular}{|c|c|c||c|c|c||c|c|c||c|c|c|} \hline
 $s$ & $h(s)$ & $p(s)$ & $s$ & $h(s)$ & $p(s)$ & $s$ & $h(s)$ & $p(s)$ 
&$s$ & $h(s)$ & $p(s)$ \\  \hline 
&&&&&&& \\[-12pt]
2 & 1& 1 &  8 & 2 & 9  &  14 & 3 & 23 & 20 & 4 & 43 \\[4pt] \hline
3 & 1& 2 &  9 & 2 & 11 &  15 & 3 & 26 & 21 & 4 & 47 \\[4pt] \hline
4 & 1& 3 &  10 & 2 & 13 & 16 & 3 & 29 & 22 & 4 & 51 \\[4pt] \hline
5 & 1& 4 &  11 & 2 & 15 & 17 & 3 & 32 & 23 & 4 & 55 \\[4pt] \hline 
6 & 2& 6 &  12 & 3 & 18 & 18 & 4 & 36 & 24 & 5 & 60 \\[4pt] \hline
7 & 1& 7 &  13 & 2 & 20 & 19 & 3 & 39 & 25 & 4 & 64 \\ \hline
\end{tabular}
\end{center}
\caption{Number of monomials $h(s)$ of $E_4$ and $E_6$ of modular
weight $2s$, and of free parameters $p(s)$ for a modular differential 
equation of order $s$.}
\end{table}

Now it is easy to see that if the equation of the form (\ref{ansatz1})
annihilates the first $k+h(s)$ powers of $q$ (starting from $q^{-k}$
up to $q^{h(s)-1}$), then the equation will hold identically, 
{\it i.e.} for each power $q^n$. The reason for this is simple: if the
differential equation annihilates the negative powers of $q$, then the
resulting function is a power series in $q$ of modular weight $2s$,
and hence must be a polynomial in $E_4$ and $E_6$ (of appropriate
degree). It is then uniquely characterised by the first $h(s)$
coefficients, starting from  $q^0$ to $q^{h(s)-1}$. If all of these
coefficients vanish, the function itself therefore has to vanish
identically.  

Thus we need to choose the order of differential equation such that
the number of free parameters, $p(s)$ is at least as big as $k+h(s)$, 
{\it i.e.}
\be
p(s) \geq k + h(s) \quad \Longleftrightarrow \quad k \leq p(s-1)  \ . 
\ee
Now consider $k=42$ and $s=21$. According to what we said above we
need to fix $k+h(s)=42+4=46$ constants, which is less then 
$p(s)=47$. Thus we expect to have an order $s=21$ modular differential
equation annihilating $\chi_{42}$, and we have checked explicitly,
using Maple, that this is indeed the case.\footnote{I thank Marco
Baumgartl for helping me do this calculation.}  
Such a differential equation should imply that
\be\label{asum}
\Phi = L_{-2}^{21} \Omega \in O_{[2]} \ .
\ee
On the other hand, $\Phi$ appears at level $42$ above the vacuum, but
for $k=42$ the proposed conformal field theory only contains Virasoro
descendants at this level. (The additional generators of the proposed
conformal field theory only appear at levels greater or equal than
$k+1=43$.) Since $c>1$ there is no null-vector relation between these
Virasoro descendants, and thus (\ref{asum}) is impossible. Thus either
our claim regarding the correspondence between the modular
differential equation and the structure of $A_{[2]}$ breaks down at
this stage, or there is a contradiction. 

Since $p(s)$ grows quadratically, it is clear that for all values of
$k\geq 42$ we obtain such a problem. (Strictly speaking, we are
assuming here, that the partition function is sufficiently generic so
that we can actually find a modular differential equation of this
order. Given that we could find such a differential equation for
$k=42$, it seems very plausible that this will also be the
case for $k\geq 43$.) For the first few $k\geq 42$ the order of the
minimal modular differential equation is summarised in table~2.   

\begin{table}[hbt]
\begin{center}
\begin{tabular}{|c|c||c|c||c|c||c|c|} \hline
$k$ & $s$ & $k$ & $s$ & $k$ & $s$ & $k$ & $s$ \\  \hline 
&&&&&&& \\[-12pt]
42 & 21 & 44 & 22 &  46 & 22 & 48 & 23  \\[4pt] \hline
43 & 21 & 45 & 22 &  47 & 22 & 49 & 23 \\ \hline
\end{tabular}
\end{center}
\caption{The minimal order $s$ of the differential equation for the
extremal self-dual theories at $c=24 k$.}
\end{table}

This analysis therefore suggests that at least the theories with 
$k\geq 42$ are inconsistent. On the other hand, by considering higher
genus amplitudes, \cite{Witten:2007kt,Gaiotto:2007xh} found impressive 
evidence that the theories with $k=2$ and $k=3$ may indeed be
consistent. The explicit genus two analysis of \cite{Gaiotto:2007xh}
can also be generalised to $k\geq 3$, but it stops being a real
consistency check at $k=11$, and thus their result is not in any
contradiction with the above suggestions.   

If these conclusions are correct, it would probably mean that one has
to adjust the chiral conformal field theory of the $AdS_3$
description, and add some (few) states at lower conformal weight. 
The conformal weight at which these additional states appear is of
order $\sqrt{k}$ for large $k$; their conformal dimension therefore
goes to infinity in the large $k$ limit, and the existence of these
states does not lead to a contradiction with the semiclassical $AdS_3$
description.\footnote{I thank Edward Witten for pointing this out to
me.} The additional states probably also do not  have any effect on
the leading order entropy calculation of \cite{Witten:2007kt}. It
would be very interesting to find a `minimal' proposal for a
consistent chiral conformal field theory satisfying these
constraints. 

That something of this nature is necessary is maybe not too surprising
in view of the superconformal analysis of \cite{Witten:2007kt}: there
it was shown that the consistency of the R-sector requires that one
has to add NS-sector states at conformal weight zero (corresponding to
$q^0$) at least for $k^*\geq 8$ (with $k^*$ even). If the above
analysis is correct, it would seem that something similar (albeit
slightly more dramatic) is required in order to make the bosonic
conformal field theories consistent.

\section*{Acknowledgements}

This research has been partially supported by the Swiss National
Science Foundation and the Marie Curie network `Constituents,
Fundamental Forces and Symmetries of the Universe'
(MRTN-CT-2004-005104). I thank Marco Baumgartl for help with the Maple
calculations, and Ilka Brunner, Terry Gannon, Peter Goddard, Sergei
Gukov, Gerald H\"ohn, Christoph Keller, Ingo Kirsch, Andy Neitzke and
Edward Witten for useful correspondences and discussions.

\end{document}